\documentclass[12pt]{article}
\usepackage{axodraw}

\catcode`@=12
\begin{document}
\thispagestyle{empty}
\begin{flushright} 
UCRHEP-T367\\ 
December 2003\
\end{flushright}
\vspace{0.5in}
\begin{center}
{\LARGE	\bf 
 Neutrinos With $Z_3$ Symmetry and\\
 New Charged-Lepton Interactions\\}
\vspace{1.5in}
{\bf Ernest Ma\\}
\vspace{0.2in}
{\sl Physics Department, University of California, Riverside, 
California 92521\\}
\vspace{1.5in}
\end{center}
\begin{abstract}\
An approximate $Z_3$ family symmetry is proposed for leptons 
which results in a neutrino mass matrix with $\sin^2 2 \theta_{atm} = 1$ 
and $\tan^2 \theta_{sol} = 0.5$, but the latter value could easily be 
smaller.  A generic requirement of this approach is the appearance of three 
Higgs doublets at the electroweak scale, resulting in possibly observable 
flavor violating leptonic decays, such as $\mu \to eee$ and $\mu \to e 
\gamma$.
\end{abstract}
\vskip 0.5in
\noindent ------------------------------

\noindent Talk at Neutrino Oscillations in Venice, December 2003.
\newpage
\baselineskip 24pt

\section{Introduction}

With the recent experimental progress in measuring atmospheric \cite{atm} 
and solar \cite{sol} neutrino oscillations, the mass-squared differences of 
the 3 active neutrinos and their mixing angles are now known with some 
precision.  In the previous talk \cite{scott}, the mixing pattern resulting 
in $\sin^2 2 \theta_{atm} = 1$ and $\tan^2 \theta_{sol} = 0.5$ was advocated. 
Here I show how this can be implemented in a complete theory of leptons, 
using the discrete family symmetry $Z_3 \times Z_2$, which is only broken 
spontaneously and by explicit soft terms in the Lagrangian.

Motivated by the idea that ${\cal M}_\nu$ should satisfy \cite{ma03,mara03}
\begin{equation}
U {\cal M}_\nu U^T = {\cal M}_\nu,
\end{equation}
where $U$ is a specific unitary matrix, a very simple form of ${\cal M}_\nu$ 
is here proposed:
\begin{equation}
{\cal M}_\nu = {\cal M}_A + {\cal M}_B + {\cal M}_C,
\end{equation}
where
\begin{equation}
{\cal M}_A = A \pmatrix {1 & 0 & 0 \cr 0 & 1 & 0 \cr 0 & 0 & 1}, ~~ 
{\cal M}_B = B \pmatrix {-1 & 0 & 0 \cr 0 & 0 & -1 \cr 0 & -1 & 0}, 
~~ {\cal M}_C = C \pmatrix {1 & 1 & 1 \cr 1 & 1 & 1 \cr 1 & 1 & 1}.
\end{equation}
This results in
\begin{equation}
\pmatrix {\nu_e \cr \nu_\mu \cr \nu_\tau} = \pmatrix {\sqrt {2/3} & 1/\sqrt 3 
& 0 \cr -1/\sqrt 6 & 1/\sqrt 3 & -1/\sqrt 2 \cr -1/\sqrt 6 & 1/\sqrt 3 & 
1/\sqrt 2} \pmatrix {\nu_1 \cr \nu_2 \cr \nu_3},
\end{equation}
with
\begin{eqnarray}
m_1 &=& A-B, \\ 
m_2 &=& A-B+3C, \\ 
m_3 &=& A+B.
\end{eqnarray}
This explains atmospheric neutrino oscillations with $\sin^2 2 \theta_{atm} 
= 1$ and solar neutrino oscillations with $\tan^2 \theta_{sol} = 1/2$. 
Whereas the mixing angles are fixed, the proposed ${\cal M}_\nu$ has the 
flexibility to accommodate the three patterns of neutrino masses often 
mentioned, i.e.

(I) the hierarchical solution, e.g. $B=A$ and $C << A$;

(II) the inverted hierarchical solution, e.g. $B = -A$ and $C << A$; 

(III) the degenerate solution, e.g. $C << B << A$.

\noindent In all cases, $C$ must be small. 

\section{Relevance of $Z_3 \times Z_2$ Symmetry}

In the above, the mxing matrix has $U_{e3} = 0$.  This is the consequence 
of a symmetry, i.e. ${\cal M}_\nu$ of Eq.~(2) is invariant under the 
$Z_2$ transformation \cite{allp}
\begin{equation}
U_2 = \pmatrix {1 & 0 & 0 \cr 0 & 0 & 1 \cr 0 & 1 & 0}, ~~~ U_2^2 = 1,
\end{equation}
However, since $C$ is always small, the possible symmetry of ${\cal M}_A + 
{\cal M}_B$ should be considered as the dominant one.  This turns out to be 
$Z_3$ \cite{z3}, i.e.
\begin{equation}
U_B = \pmatrix {-1/2 & -\sqrt{3/8} & -\sqrt{3/8} \cr \sqrt{3/8} & 1/4 & 
-3/4 \cr \sqrt{3/8} & -3/4 & 1/4}, ~~~ U_B^3 = 1. 
\end{equation}
Note that $U_B$ commutes with $U_2$ and ${\cal M}_\nu = {\cal M}_A + 
{\cal M}_B$ is the most general solution of
\begin{equation}
U_B {\cal M}_\nu U_B^T = {\cal M}_\nu.
\end{equation}

\section{Origin of ${\cal M}_C$}

Since ${\cal M}_C$ is small and breaks the symmetry of ${\cal M}_A + 
{\cal M}_B$, it is natural to think of its origin in terms of the well-known 
dimension-five operator \cite{w79}
\begin{equation}
{\cal L}_{eff} = {f_{ij} \over 2 \Lambda} (\nu_i \phi^0 - l_i \phi^+)(\nu_j 
\phi^0 - l_j \phi^+) + H.c.,
\end{equation}
where $(\phi^+,\phi^0)$ is the usual Higgs doublet of the Standard Model and 
$\Lambda$ is a very high scale.  As $\phi^0$ picks up a nonzero vacuum 
expectation value $v$, neutrino masses are generated, and if $f_{ij} v^2/
\Lambda = C$ for all $i,j$, ${\cal M}_C$ is obtained.  Since $\Lambda$ is 
presumably of order $10^{16}$ to $10^{18}$ GeV, $C$ is of order $10^{-3}$ 
to $10^{-5}$ eV.  This range of values is just right to encompass all 
three solutions mentioned above.

As for the form of ${\cal M}_C$, it may be understood as coming from effective 
universal interactions among the leptons at the scale $\Lambda$.  For example, 
if Eq.~(11) has an $S_3$ symmetry as generated by $U_2$ and \cite{z3}
\begin{equation}
U_C = \pmatrix {0 & 1 & 0 \cr 0 & 0 & 1 \cr 1 & 0 & 0}, ~~~ U_C^3 = 1, 
\end{equation}
the most general form of ${\cal M}_C$ would be
\begin{equation}
{\cal M}_C = C \pmatrix {1 & 1 & 1 \cr 1 & 1 & 1 \cr 1 & 1 & 1} + 
C' \pmatrix {1 & 0 & 0 \cr 0 & 1 & 0 \cr 0 & 0 & 1}.
\end{equation}
However, the $C'$ term can be absorbed into ${\cal M}_A$, so again 
${\cal M}_\nu$ of Eq.~(2) is obtained.  This form of the neutrino mass 
matrix has in fact been discussed as an ansatz in a number of recent 
papers \cite{hps02,xing,hs02,hs03,hz}.  In particular, let it be 
rewritten as
\begin{equation}
{\cal M}_\nu = (A+C) \pmatrix {1 & 0 & 0 \cr 0 & 1 & 0 \cr 0 & 0 & 1} - 
B \pmatrix {1 & 0 & 0 \cr 0 & 0 & 1 \cr 0 & 1 & 0} + C \pmatrix {0 & 1 & 0 \cr 
0 & 0 & 1 \cr 1 & 0 & 0} + C \pmatrix {0 & 0 & 1 \cr 1 & 0 & 0 \cr 0 & 1 & 0}.
\end{equation}
Note that each of the above four matrices is a group element of $S_3$.  This 
is the recent proposal of Harrison and Scott \cite{hs03}.  The difference 
here is that the underlying symmetry of ${\cal M}_\nu$ has been identified, 
thus allowing a complete theory of leptons to be constructed.

\section{Origin of ${\cal M}_A + {\cal M}_B$}

To accommodate the proposed $Z_3$ symmetry in a complete theory, the 
Standard Model of particle interactions is now extended 
\cite{mara03} to include three scalar doublets $(\phi^0_i, \phi^-_i)$ 
and one very heavy triplet $(\xi^{++}, \xi^+, \xi^0)$.  The leptonic Yukawa 
Lagrangian is given by
\begin{eqnarray}
{\cal L}_Y = h_{ij} [\xi^0 \nu_i \nu_j - \xi^+ (\nu_i l_j + l_i \nu_j)/
\sqrt 2 + \xi^{++} l_i l_j] 
+ f_{ij}^k (l_i \phi^0_j - \nu_i \phi^-_j) l^c_k + H.c.,
\end{eqnarray}
where, under the $Z_3$ transformation,
\begin{eqnarray}
&& (\nu,l)_i \to (U_B)_{ij} (\nu,l)_j, ~~~ l^c_k \to l^c_k, \\ 
&& (\phi^0,\phi^-)_i \to (U_B)_{ij} (\phi^0,\phi^-)_j, ~~~ (\xi^{++}, \xi^+, 
\xi^0) \to (\xi^{++}, \xi^+, \xi^0).
\end{eqnarray}
This means
\begin{equation}
U_B^T h U_B = h, ~~~ U_B^T f^k U_B = f^k,
\end{equation}
resulting in
\begin{equation}
h = \pmatrix {a-b & 0 & 0 \cr 0 & a & -b \cr 0 & -b & a}, ~~~ f^k = \pmatrix 
{a_k - b_k & d_k & d_k \cr -d_k & a_k & -b_k \cr -d_k & -b_k & a_k}.
\end{equation}
Note that $h$ has no $d$ terms because it has to be symmetric.  Note also 
that both $h$ and $f$ are invariant under $U_2$ of Eq.~(8). 
The neutrino mass matrix ${\cal M}_A + {\cal M}_B$ is obtained with $A = 2a 
\langle \xi^0 \rangle$ and $B = 2b \langle \xi^0 \rangle$.  The fact that 
it is proportional to a \underline {single} vacuum expectation value is 
important for the preservation of the $Z_3$ symmetry.  As for the 
smallness of $\langle \xi^0 \rangle$, it is fully explained 
\cite{masa98,ma98} as the analog of the canonical seesaw mechanism in the 
case of very large and positive $m_\xi^2$.

The $Z_3$ symmetry is  
broken by the soft terms of the Higgs sector, thus $v_{1,2} << v_3$ may be 
assumed. If $d_k,b_k << a_k$ is also assumed (which by itself does not break 
$Z_3$), then the hierarchy of $m_e, m_\mu, m_\tau$ is understood. 
The mixing matrix $V_L$ in the $l_i$ basis (such that 
$V_L {\cal M}_l {\cal M}_l^\dagger V_L^\dagger$ 
is diagonal) will be very close to the identity matrix with off-diagonal 
terms of order $m_e/m_\mu$, $m_e/m_\tau$, and $m_\mu/m_\tau$.  This 
construction allows ${\cal M}_\nu$ of Eq.~(2) to be in the $(\nu_e, \nu_\mu, 
\nu_\tau)$ basis as a very good approximation.

\section{Flavor Violating Leptonic Decays}

The Yukawa couplings of the three Higgs doublets are such that the dominant 
coupling of $\phi_1^0$ is $(m_\tau/v_3) e \tau^c$ and that of 
$\phi_2^0$ is $(m_\tau/v_3)\mu \tau^c$.  Other couplings are at most of 
order $m_\mu/v_3$ in this model, and some are only of order $m_e/v_3$.  
The smallness of flavor changing decays in the leptonic sector is thus 
guaranteed, even though they should be present and may be observable in 
the future.  The decays $\tau^- \to e^- e^+ e^-$ and $\tau^- 
\to e^- e^+ \mu^-$ may proceed through $\phi_1^0$ exchange with coupling 
strengths of order $m_\mu m_\tau/v_3^2 \simeq (g^2/2) (m_\mu m_\tau/M_W^2)$, 
whereas the decays $\tau^- \to \mu^- \mu^+ \mu^-$ and $\tau^- \to \mu^- \mu^+ 
e^-$ may proceed through $\phi_2^0$ exchange also with coupling strengths of 
the same order.  We estimate the order of magnitude of these branching 
fractions to be
\begin{equation}
B \sim \left( {m_\mu^2 m_\tau^2 \over m_{1,2}^4} \right) B(\tau \to \mu \nu 
\nu) \simeq 6.1 \times 10^{-11} \left( {100~{\rm GeV} \over m_{1,2}} \right)^4,
\end{equation}
which is well below the present experimental upper bound of the order 
$10^{-6}$ for all such rare decays \cite{pdg}.

The decay $\mu^- \to e^- e^+ e^-$ may also proceed through $\phi_1^0$ with 
a coupling strength of order $m_\mu^2/v_3^2$.  Thus
\begin{equation}
B(\mu \to e e e) \sim {m_\mu^4 \over m_1^4} \simeq 1.2 \times 10^{-12} \left( 
{100~{\rm GeV} \over m_1} \right)^4,
\end{equation}
which is at the level of the present experimental upper bound of $1.0 \times 
10^{-12}$.  The decay $\mu \to e \gamma$ may also proceed through $\phi_2^0$ 
exchange (provided that $Re \phi_2^0$ and $Im \phi_2^0$ have different masses) 
with a coupling of order $m_\mu m_\tau/v_3^2$.  Its branching fraction is 
given by \cite{mara01}
\begin{equation}
B(\mu \to e \gamma) \sim {3 \alpha \over 8 \pi} {m_\tau^4 \over m_{eff}^4},
\end{equation}
where
\begin{equation}
{1 \over m_{eff}^2} = {1 \over m_{2R}^2} \left( \ln {m_{2R}^2 \over m_\tau^2} 
- {3 \over 2} \right) - {1 \over m_{2I}^2} \left( \ln {m_{2I}^2 \over 
m_\tau^2} - {3 \over 2} \right).
\end{equation}
Using the experimental upper bound \cite{meg} of $1.2 \times 10^{-11}$, we 
find $m_{eff} > 164$ GeV.

Once $\phi_1^0$ or $\phi_2^0$ is produced, its dominant decay will be to 
$\tau^\pm e^\mp$ or $\tau^\pm \mu^\mp$ if each couples only to leptons. 
If they also couple to quarks (and are sufficiently heavy), then the dominant 
decay products will be $t \bar u$ or $t \bar c$ together with their 
conjugates.  As for $\phi_3^0$, it will behave very much as the single 
Higgs doublet of the Standard Model, with mostly diagonal couplings to 
fermions.  It should also be identified with the $\phi$ of Eq.~(15).

\section{Consequences of an Arbitrary ${\cal M}_C$}

It should also be noted that as long as 
${\cal M}_C$ is small, its exact form is not that important.  Let
\begin{equation}
{\cal M}_C = \pmatrix {a & d & e \cr d & b & f \cr e & f & c},
\end{equation}
then it is easily shown that
\begin{equation}
\tan^2 \theta_{sol} = \left( {1 - \sqrt {1+z^2} \over z} \right)^2,
\end{equation}
where
\begin{equation}
z = {2 \sqrt 2 (d+e) \over 2(f-a) + b + c}.
\end{equation}
For $z = 2 \sqrt 2$, $\tan^2 \theta_{sol} = 0.5$ is recovered.  If $z=2.2$ 
instead, $\tan^2 \theta_{sol} = 0.42$ is obtained, which is the central value 
of this parameter in a recent global fit \cite{fit} of all the data.  Also, 
$U_{e3}$ becomes nonzero in general and is given by
\begin{equation}
U_{e3} \simeq {d-e \over 2 \sqrt 2 B}.
\end{equation}

\section{Conclusion}

If a symmetry is indeed responsible for the observed pattern of neutrino 
masses and mixing, then $Z_3 \times Z_2$ is a very good candidate.  It 
explains the dominant structure of ${\cal M}_\nu$, i.e. ${\cal M}_A$ and 
${\cal M}_B$ of Eq.~(3), which are proportional to the $vev$ of a single 
Higgs triplet.  The remainder, i.e. ${\cal M}_C$ of Eq.~(3) or (24), should 
be considered as a perturbation from another source of neutrino mass, such 
as the effective dimension-five operator of Eq.~(11).  Arbitrary 
charged-lepton mass may be accommodated and flavor violating leptonic 
decays such as $\mu \to eee$ and $\nu \to e \gamma$ are predicted to be in 
the observable range.

\section{Acknowledgements}

I thank Milla Baldo Ceolin for her great hospitailty in Venice.  This work was 
supported in part by the U.~S.~Department of Energy
under Grant No.~DE-FG03-94ER40837.


\begin{thebibliography}{99}
\bibliographystyle{unsrt}
\bibitem{atm} C. K. Jung, C. McGrew, T. Kajita, and T. Mann, Ann. Rev. Nucl. 
Part. Sci. {\bf 51}, 451 (2001).
\bibitem{sol} Q. R. Ahmad {\it et al.}, SNO Collaboration, Phys. Rev. Lett. 
{\bf 89}, 011301, 011302 (2002); nucl-ex/0309004; K. Eguchi {\it et al.}, 
KamLAND Collaboration, Phys. Rev. Lett. {\bf 90}, 021802 (2003).
\bibitem{scott} W. G. Scott, these proceedings.
\bibitem{ma03} E. Ma, Phys. Rev. Lett. {\bf 90}, 221802 (2003).
\bibitem{mara03} E. Ma and G. Rajasekaran, Phys. Rev. {\bf D68}, 071302(R) 
(2003).
\bibitem{allp} E. Ma, Phys. Rev. {\bf D66}, 117301 (2002).
\bibitem{z3} E. Ma, hep-ph/0308282.
\bibitem{w79} S. Weinberg, Phys. Rev. Lett. {\bf 43}, 1566 (1979).
\bibitem{hps02} P. F. Harrison, D. H. Perkins, and W. G. Scott, Phys. Lett. 
{\bf B530}, 167 (2002).
\bibitem{xing} Z.-Z. Xing, Phys. Lett. {\bf B533}, 85 (2002).
\bibitem{hs02} P. F. Harrison and W. G. Scott, Phys. Lett. {\bf B535}, 163 
(2002).
\bibitem{hs03} P. F. Harrison and W. G. Scott, Phys. Lett. {\bf B557}, 76 
(2003).
\bibitem{hz} X.-G. He and A. Zee, Phys. Lett. {\bf B560}, 87 (2003).
\bibitem{masa98} E. Ma and U. Sarkar, Phys. Rev. Lett. {\bf 80}, 5716 (1998).
\bibitem{ma98} E. Ma, Phys. Rev. Lett. {\bf 81}, 1171 (1998).
\bibitem{pdg} K. Hagiwara {\it et al.}, Particle Data Group, Phys. Rev. 
{\bf D66}, 011501 (2002).
\bibitem{mara01} E. Ma and G. Rajasekaran, Phys. Rev. {\bf D64}, 113012 (2001).
\bibitem{meg} M. L. Brooks {\it et al.}, Phys. Rev. Lett. {\bf 83}, 1521 
(1999).
\bibitem{fit} P. Aliani {\it et al.}, hep-ph/0309156.
\end{thebibliography}
\end{document}